\documentclass[epj,twocolumn]{webofc}
\usepackage[varg]{txfonts}   
\usepackage{hyperref}
\woctitle{ISVHECRI 2018} 
\begin{document}
\raggedbottom
%
\title{Astrophysical neutrino production and impact of associated uncertainties \\ in photo-hadronic interactions of UHECRs}
%
%

\author{\firstname{Daniel} \lastname{Biehl}\inst{1}\fnsep\thanks{Speaker, \email{daniel.biehl@desy.de}}
      \and \firstname{Denise} \lastname{Boncioli}\inst{1}\fnsep
      \and \firstname{Anatoli} \lastname{Fedynitch}\inst{1}\fnsep
      \and \firstname{Leonel} \lastname{Morejon}\inst{1}\fnsep
      \and \firstname{Walter} \lastname{Winter}\inst{1}\fnsep
}

\institute{Deutsches Elektronen-Synchrotron (DESY), Platanenallee 6, 15738 Zeuthen, Germany
          }

\abstract{%
  High energy neutrinos can be produced by interactions of ultra-high energy cosmic rays (UHECRs) in the dense radiation fields of their sources as well as off the cosmic backgrounds when they propagate through the universe. Multi-messenger interpretations of current measurements deeply rely on the understanding of these interactions. In order to efficiently produce neutrinos in the sources of UHECRs, at least a moderate level of interactions is needed, which means that a nuclear cascade develops if nuclei are involved. On the other hand, the available cross-section data and interaction models turn out to make poor predictions for most nuclei heavier than protons. We show the impact of these uncertainties in state-of-the-art photo-disintegration models and motivate nuclear cross-section measurements. Further, we discuss extensions for photo-meson models currently used in astrophysics and demonstrate the importance of understanding the details of UHECR interaction with the Glashow resonance.
}
%


%
\maketitle


\section{Introduction}
More than 100 years after their discovery, the origin of ultra-high energy cosmic rays (UHECRs) is still a mystery. Multi-messenger astronomy provides powerful tools to identify their sources by connecting air-shower observations with astrophysical neutrino, gamma-ray and now also gravitational wave measurements. In order to interpret the connection among the different messengers, it is essential to understand cosmic ray interactions in the sources, during propagation as well as in the atmosphere on Earth.

Some of the most powerful sources known to date like Gamma-Ray Bursts (GRBs) \cite{Piran:2004ba}, Active Galactic Nuclei (AGNs) \cite{Stecker:1991vm} or Tidal Disruption Events (TDEs) \cite{Wang:2011ip,Murase:2013ffa} are promising sites to produce astrophysical neutrinos in UHECR interactions. The dominant processes for neutrino production in such sources are of photo-hadronic nature, i.e. baryons interacting with the ambient photon fields for example via the $\Delta$-resonance and, depending on the source characteristics, subsequent meson decays.

The maximum energy charged particles can be accelerated to depends mainly on the magnetic field and the size of the accelerator.
Cosmic accelerators easily reach sizes of billions of kilometres, such that the maximum energy of cosmic rays can reach up to about $10^{21}$ eV. Target photons with comparably low energy interact with these ultra-high energy particles, triggering photo-nuclear interactions in the MeV -- GeV range (as seen in the rest frame of the nucleus).

It turns out that the available experimental data necessary to build reliable models are sparse as cross-sections are measured only for a few isotopes \cite{Otuka:2014wzu}. For all other isotopes, model predictions are needed, which are not always reproducing the data. Depending on how many and which isotopes are included in the disintegration chain of a nucleus, the results can therefore change drastically.

Here, we investigate the model dependence of neutrino production in UHECR interactions. We discuss the impact on disintegration rates, neutrino fluxes as well as on the flavor composition, which is especially important in future generation neutrino telescopes such as IceCube-Gen2. Further we discuss possible improvements and give a motivation for conducting more nuclear cross-section measurements for models in astrophysics.


\section{Energy scales of photo-nuclear interactions}

The discussion presented in the following is valid for photo-hadronic interactions in arbitrary environments. However, a possible, generic picture for sources dominated by photo-hadronic interactions is given by a central engine which launches relativistic shells with different Lorentz factors. These shells will collide with each other and produce shocks, where particles are accelerated and interactions are triggered.

Photo-hadronic ($A\gamma$-) interactions happen on different energy scales. A possible classification in terms of the photon energy in the nucleus rest frame $\varepsilon_r$ is given by
\begin{itemize}
\item \textbf{QED scale:} e.g. electon-positron pair production
	\begin{equation} A + \gamma \rightarrow A + e^+ + e^-, \quad \varepsilon_r > 1 \text{ MeV} \end{equation}
\item \textbf{Nuclear scale:} nuclear photo-disintegration, e.g.
	\begin{equation} A + \gamma \rightarrow (A-1) + n, \quad \varepsilon_r > 8 \text{ MeV} \end{equation}
\item \textbf{Mesonic scale:} baryonic resonances, photo-meson production (producing neutrinos), e.g.
	\begin{equation} A + \gamma \rightarrow \tilde{A} + \pi^+, \quad \varepsilon_r > 140 \text{ MeV} \end{equation}
\item \textbf{Hadronic scale:} hadronic structure becomes relevant for the interaction ($\varepsilon_r > 1$ GeV)
\end{itemize}

Interactions on the QED scale, with energy losses due to pair production as a prominent example, is highly relevant and has been intensively studied in the context of cosmic ray propagation in interactions with the cosmic microwave background (CMB). In the above mentioned source classes of UHECRs, the nuclear and mesonic scale are typically the dominant processes given that the radiation density is not too low \cite{Biehl:2017zlw,Rodrigues:2017fmu,Biehl:2017hnb}. For the composition of cosmic rays, nuclear disintegration is the main influence while for the neutrino production, photo-meson processes are most important. The hadronic scale becomes relevant only for very high photon energies (in the nucleus rest frame), i.e. it does not affect the peak neutrino flux produced by baryonic resonances very much.

Let us note for the sake of completeness that there are many other processes, such as beta decays, pp-/AA-interactions, spontaneous nucleon emission, spallation, de-excitation, etc. However, here we focus on the most relevant ones for UHECR interactions and neutrino production, which is photo-disintegration and photo-meson production, respectively.


\section{Development of the nuclear cascade}

In theory, a nuclear disintegration chain can be calculated fully deterministic by solving a system of coupled transport equations 
\begin{equation}
\frac{\partial N_i}{\partial t} = \frac{\partial}{\partial E} (-b(E)N_i(E)) - \frac{N_i(E)}{t_\text{esc}} + \tilde{Q}_{ji}(E)
\end{equation}
for several particle species $i$ (e.g. a nuclear isotope). The first term on the right hand side corresponds to energy losses $b(E) = E t^{-1}_\text{loss}$ with the energy loss rate $t^{-1}_\text{loss}$. The second term represents particle escape with a rate $t^{-1}_\text{esc}$. The coupling of different species is due to the term
\begin{equation}
\tilde{Q}_{ji}(E) = Q_i(E) + Q_{j \rightarrow i}(E),
\end{equation}
which contains an injection term $Q_i(E)$ from an acceleration zone and an injection term $Q_{j \rightarrow i}(E)$ from other species, such as from disintegration. The system of partial differential equations (PDEs) is to be solved for differential densities $N_i$ [GeV$^{-1}$ cm$^{-3}$] in the shock rest frame (SRF).

\begin{figure}[h]
\centering
\includegraphics[width=0.234\textwidth]{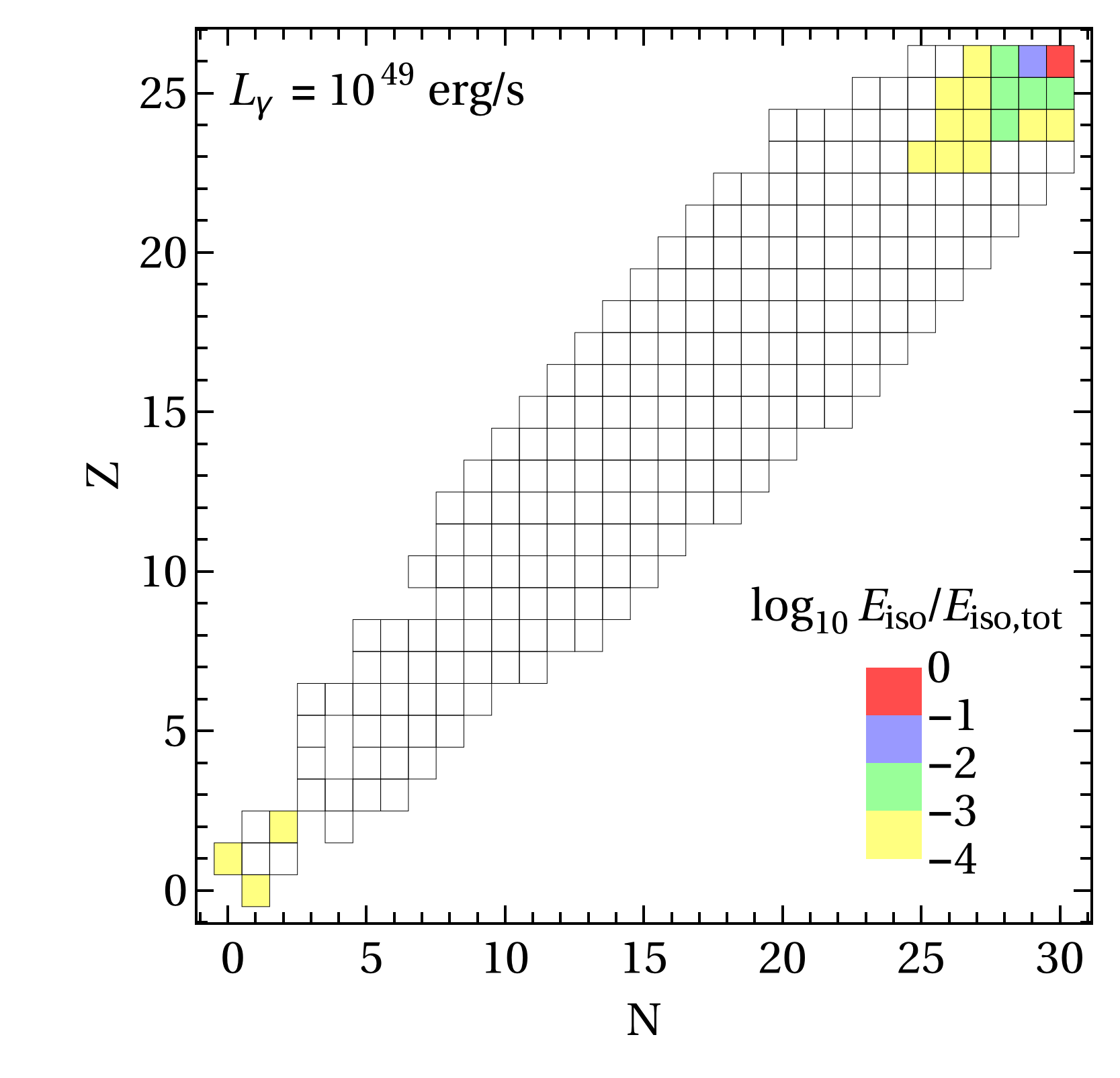}
\includegraphics[width=0.234\textwidth]{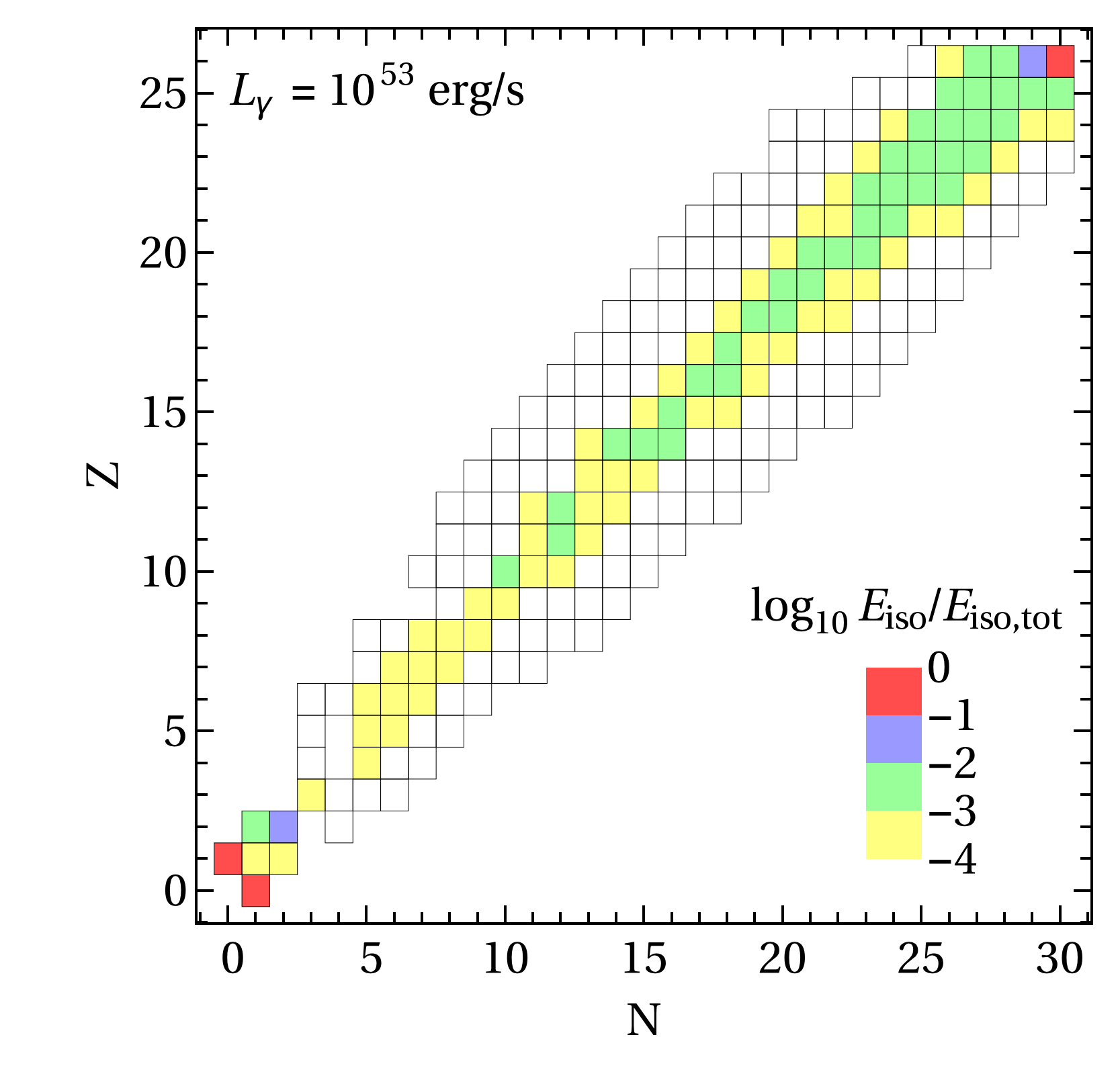}
\caption{Isotope chart depicting the development of the nuclear cascade in the source for different luminosities and pure $^{56}$Fe injection. The color code indicates the energy per isotope relative to the total isotropic equivalent energy.
Note that the size of the region is kept constant. Figure taken from \cite{Biehl:2017zlw}.}
\label{fig:nuclearcascade}
\end{figure}

A schematic representation of the time evolution of such a system is the so-called nuclear cascade shown in Fig.~\ref{fig:nuclearcascade}. Every square in the nuclear chart represents one isotope species $i$ in the PDE with charge number $Z$ and neutron number $N$. The different species are connected by many competing channels representing the coupling (not shown). Here, the illustration depicts the example of pure $^{56}$Fe injected in a GRB shell for different luminosities. Depending on the parameters, different species will be populated as a consequence of the interactions as indicated by the color code, which gives the energy per isotope relative to the total isotropic equivalent energy.

The development of the nuclear cascade as well as the efficiency of neutrino production scales with the radiation density of the interaction region $u_\gamma \propto L_\gamma/R^2$. If we place ourselves in the internal shock scenario with $R \approx 2\Gamma^2c t_v$, we recover the well known scaling relation for the pion production efficiency
$f_{\pi} \propto L_\gamma /(\Gamma^4 t_v)$ \cite{Waxman:1998yy,Guetta:2003wi}. Thus, the size of the region $R$ and gamma-ray luminosity $L_\gamma$ are the main control parameters for the nuclear cascade and neutrino production. Note that in Fig.~\ref{fig:nuclearcascade} the size of the region is kept constant for both panels. As a consequence of the increased radiation density, many more isotopes are populated and a significant fraction of the energy is efficiently dumped into nucleons. For more details, see \cite{Biehl:2017zlw}.


\section{Cross-section data and models}
\label{sec:4-0}
\begin{figure}
\centering
\includegraphics[width=0.34\textwidth]{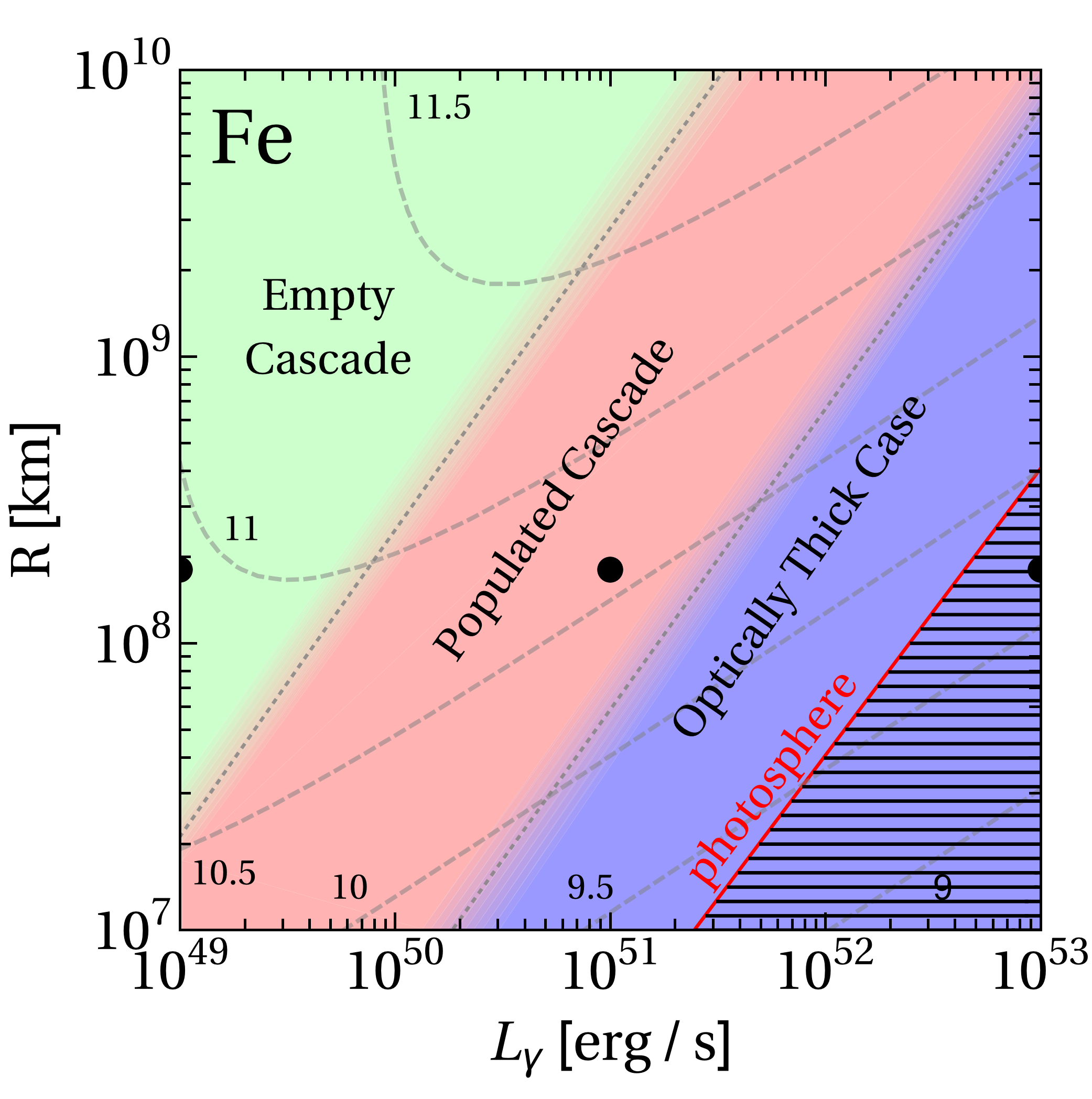}
\caption{Parameter space scan in gamma-ray luminosiy $L_\gamma$ and size of the region $R$ showing the three different regimes (see main text for details) of the nuclear cascade (pure $^{56}$Fe injected). The transitions between the regimes are indicated by the gray dotted lines. Sub-photospheric emission is obtained for parameters in the hatched region. The primary maximum energy in the observer's frame (in GeV) is superimposed by the gray contours and the black dots represent the parameters for the nuclear cascade shown in Fig.~\ref{fig:nuclearcascade} ($L_\gamma = 10^{51}$ erg s$^{-1}$ not shown). Note that the Lorentz factor is fixed to $\Gamma \sim 300$. Figure taken from \cite{Biehl:2017zlw}.
}
\label{fig:parameterspace}
\end{figure}

With the nuclear cascade as a measure of cosmic ray interactions and neutrino production, the parameter space in luminosity and size of the region can be divided into regions of no, moderate and efficient interactions. The result of this scan is illustrated in Fig.~\ref{fig:parameterspace} for the GRB example mentioned above. Note that this procedure is, in principle, applicable to arbitrary interaction zones.

Fig.~\ref{fig:parameterspace} shows three different regions. For low radiation densities (upper left corner) we are in the Empty Cascade (green) region, where the nuclear cascade does not fully develop and typically only a few elements close to the injected composition are populated. This is due to the low interaction rates, meaning that the radiation zone is optically thin to photo-hadronic interactions of all cosmic ray species. In this case, the neutrino fluxes are very low as a result of the low radiation densities. With increasing radiation density (going towards the lower right corner), we encounter the Populated Cascade (red) and the source starts to become optically thick to photo-hadronic interactions of the heaviest mass nuclei (but remains transparent to nucleons), because the photo-hadronic cross-section scales with $A$. The cascade can develop broadly along the main diagonal along with an intermediate neutrino production. High radiation densities lead to the Optically Thick Case, where the source is opaque to photo-hadronic interactions of all species. A narrow cascade develops with very efficient production of nucleons and neutrinos.

The transitions between the different regions are indicated by the gray dotted lines and the maximum reachable energy (in the observer's frame in GeV) is depicted by the gray dashed contours. The radiation density can even be too high for photons to escape, which is indicated by the hatched region. The photospheric radius becomes larger than the emission radius for these parameter combinations, such that strictly speaking this model is not valid anymore because the target photon spectrum changes. Note that in this example, the Lorentz factor is fixed to $\Gamma \sim 300$.

While the neutrino production in the Optically Thick Case is dominated by photo-hadronic interactions of nucleons, neutrinos in the Empty Cascade region originate mostly from interactions of heavy nuclei. The important difference is that $p\gamma / n\gamma$-interactions are well understood \cite{Mucke:1999yb} compared to their heavy counterparts. The cross-sections for neutrino production in the Empty and Populated Cascade regions are highly depending on nuclear interaction models, which are not always describing their behaviour very well. This introduces uncertainties, such that the neutrino flux predictions in environments with lower radiation densities are in principle less reliable due to the lack of nuclear data. In fact, the available cross-section data are very sparse \cite{Otuka:2014wzu}, highlighting the need of future measurements and improved theoretical models.


\subsection{Uncertainties from photo-disintegration}

\begin{figure}[t!]
\centering
\includegraphics[width=0.4\textwidth]{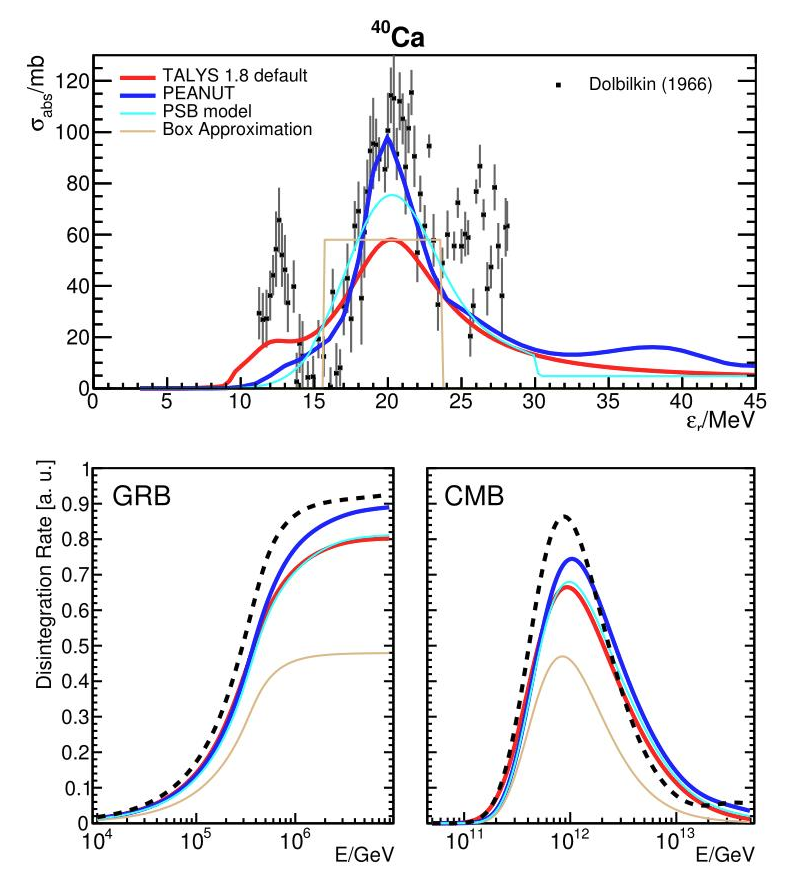}
\caption{Cross section data (top) and disintegration rates in the radiation field of a GRB (bottom left) as well as in the CMB (bottom right) for the double-magic nucleus $^{40}$Ca. The GRB spectrum is described by a broken power law with spectral index $\alpha = 1.0$ below and $\beta = 2.0$ above the break at 1 keV (SRF), while the obtained CMB rate assumes a thermal target photon field with $T = 2.73$ K. Figure taken from \cite{Boncioli:2016lkt}.}
\label{fig:crosssection}
\end{figure}

In astrophysical sources as GRBs, photo-disintegration can be the dominant energy loss process for nuclei. The in-source composition of cosmic rays can be very important for neutrino production as discussed in the last paragraph of Sec.~\ref{sec:4-0}. The lack of data requires (at least some) more measurements with which one can build a reliable model. However, current models do not always describe the photon absorption very well \cite{Batista:2015mea,Boncioli:2016lkt,Soriano:2018lly}. Fig.~\ref{fig:crosssection} shows the measured cross-section of $^{40}$Ca (black data points) as well as certain cross-section models in the top panel and the respective predictions for disintegration rates in the radiation field of a GRB (left) and in the CMB (right) in the bottom panels. The box approximation (brown curve), as for instance defined in \cite{Murase:2010gj}, underestimates data and models. This leads to up to a factor two difference in disintegration rates compared to data and other models, which is clearly an insufficient description. 

$^{40}$Ca is a very special nucleus as it is double-magic. Predictions of the TALYS \cite{Koning:2007} (CRpropa \cite{Kampert:2012fi}) model however depend only weakly on the nuclear mass and element, such that the predicted cross-section for $^{40}$Ar shows the same behaviour. Although no measurements are available for $^{40}$Ar, due to the different shell structure it is expected to have a different cross-section. PEANUT (which is a model of FLUKA) \cite{Ferrari:2005zk} predictions are different in the same isobar. For the isotopes for which measurements are available, at least the central peak of the giant dipole resonance (GDR) is reproduced. 

\begin{figure}[ht]
\centering
\includegraphics[width=0.5\textwidth]{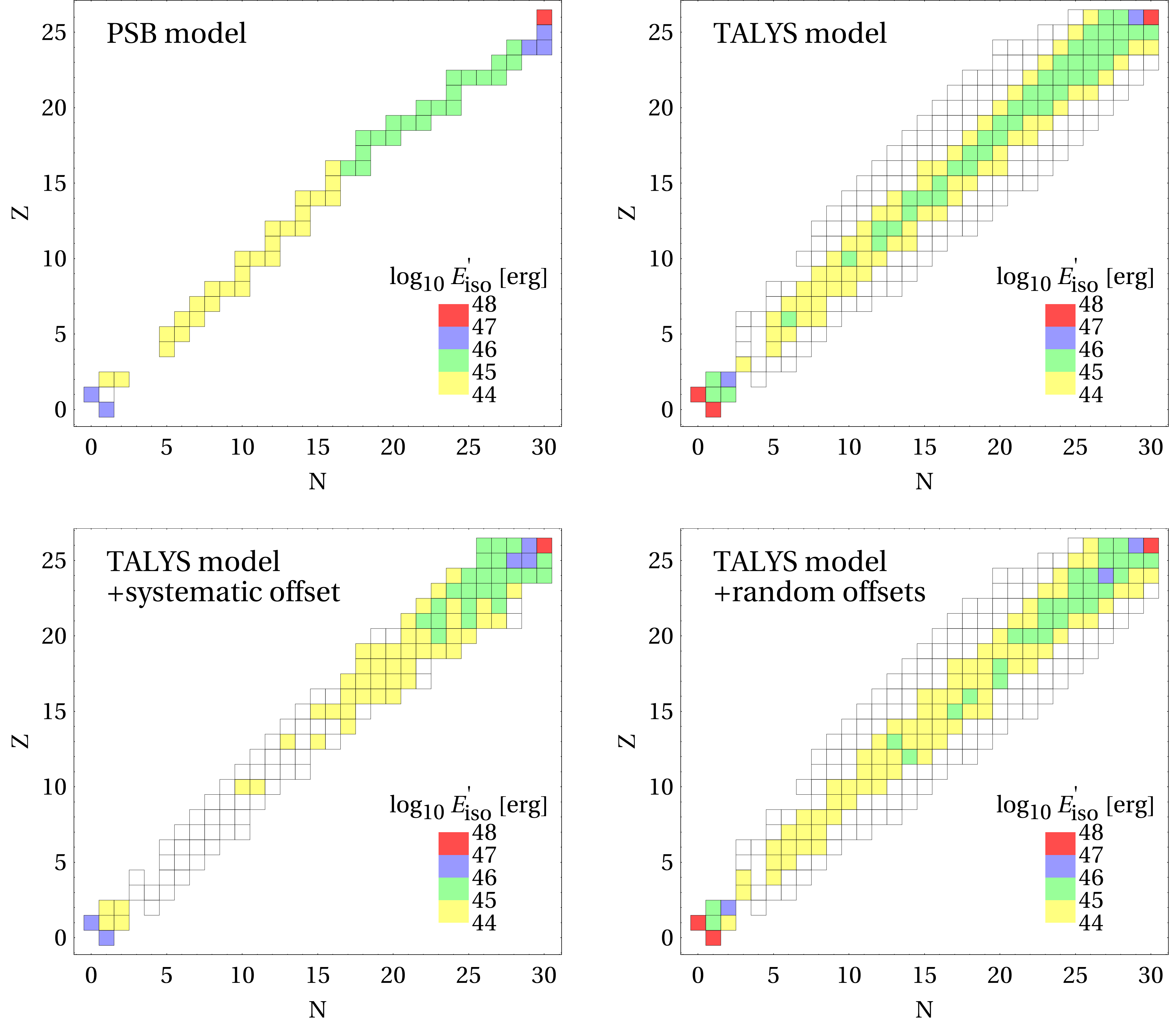}
\caption{Disintegration chain of $^{56}$Fe in a GRB shell with the colors representing the total energy per isotope in the SRF (see legend). White boxes correspond to energies lower than $10^{44}$ erg s$^{-1}$. The upper row shows the PSB model (left) and TALYS (right, CRPropa2 for the lighter elements), while the lower row shows the TALYS setup including different systematic assumptions (see main text). The GRB parameters are $L_{\gamma,\text{iso}} = 10^{52}$ erg s$^{-1}$, $\Gamma = 300$, $t_v = 0.01$ s, $z = 2$ and an $E^{-2}$ injection spectrum. Figure taken from \cite{Boncioli:2016lkt}.}
\label{fig:interactionmodels}
\end{figure}

Different interaction models have a direct impact on the nuclear cascade, which is shown in Fig.~\ref{fig:interactionmodels} for the example of pure iron injection in a GRB shell. The PSB model \cite{Puget:1976nz} (upper left) has a one-dimensional disintegration chain and because they are ejected in each interaction, protons and neutrons are strongly populated. On the other hand, TALYS (upper right) provides much more channels, such as the ejection of protons, neutrons, deuterium, tritium, He-3 or He-4. Also elements off the main diagonal are populated and significantly more light to intermediate elements are produced. In the bottom panels, different systematic assumptions for the TALYS model are shown. The 'systematic offset' case (lower left) is obtained by choosing the cross-section to be 0 for all unmeasured, and 0.5 of their nominal values for the cases in which only inclusive channels are measured. Because of this suppression of the cross-section, the nuclear cascade will cease to develop. Light and intermediate nuclei will not be populated. The 'random offset' case varies unmeasured cross-sections randomly between 0 and 2 of their nominal values. Partially measured cross-sections will vary randomly between 0.5 and 1.5. With this systematics assumption, the total cascade looks similar to the TALYS one, but certain isotopes will be populated differently. The choice of the interaction model directly influences the escaping cosmic ray composition.


\subsection{Uncertainties in photo-meson production}

Current state-of-the-art photo-meson models in astrophysics often parameterize the cross-section of a nucleus as a superposition of protons and neutrons
\begin{equation}
\sigma^\text{tot}_A(E) = \frac{Z}{A} \sigma_p(E) + \frac{N}{A} \sigma_n(E),
\end{equation}
assuming individual nucleon interaction. After the interaction, one nucleon is kicked out from the nucleus while the remaining fragment is re-injected into the system without any mediating de-excitations or decays. We currently test the impact of these effects on UHECR and neutrino production in astrophysical objects.

\begin{figure}[t!]
\centering
\includegraphics[width=0.34\textwidth]{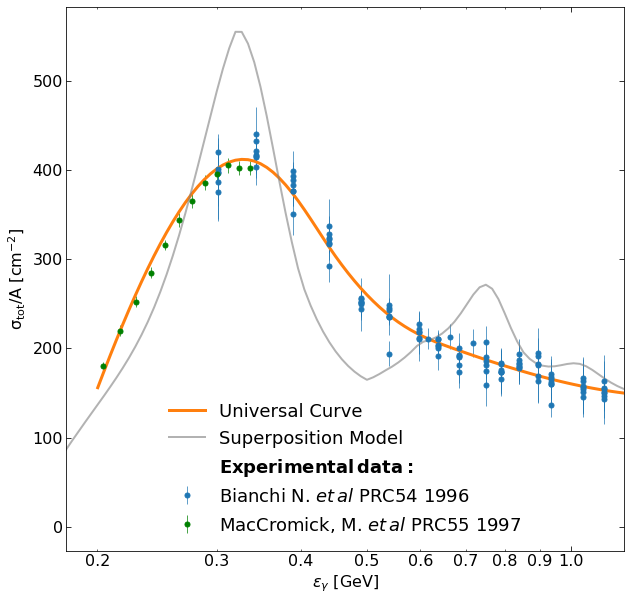}
\caption{Cross-section over mass number as a function of the photon-energy in the nucleus rest frame. The data points show measurements of different isotopes, suggesting a universal cross-section behaviour for all nuclei heavier than protons (orange curve). The gray curve represents the superposition model, which approximates the nucleus' cross-section by $\sigma_{A\gamma} = \frac{Z}{A}\sigma_{p\gamma} + \frac{N}{A}\sigma_{n\gamma}$ with the proton and neutron cross-sections $\sigma_{p\gamma}$ and $\sigma_{n\gamma}$. Figure taken from \cite{Morejon:2018}.}
\label{fig:universalcurve}
\end{figure}

Rather than a superposition of nucleons, a universal cross-section behaviour for photo-meson production is suggested by data. In Fig.~\ref{fig:universalcurve}, the total cross-section divided by the mass number is shown as a function of the photon energy in the nucleus rest frame. The gray curve indicates the prediction of the superposition model, while the data aligns along the orange curve, which we call universal curve. A possible explanation for this behaviour is that the interaction does not happen in an isolated fashion with one nucleon only, but rather there are re-absorption and emission processes of the nearby nucleons, which makes the curve smooth out. The cross-section scales with the mass number $A$, in the energy range from $\sim 140$ MeV to $\sim 1$ GeV. Above these energies, the scaling changes according to the Glauber rule $\propto A^{2/3}$. However, some models assume this scaling for the whole energy range.

Furthermore, we investigate the influence of multi-nucleon emission in photo-meson models. To do so, we define our toy model as the nuclear cascade of a TDE with pure nitrogen emission, motivated by the tidal disruption of carbon-oxygen white dwarfs \cite{Biehl:2017hnb}. This is suitable to explore the mesonic scale since the photo-disintegration in TDEs can be suppressed with respect to the GRBs, depending on spectral shape of the target photon field. In this scenario, more energy is dumped in specific channels along the main diagonal, making heavy nuclei disintegrate faster because intermediate steps can be skipped. Especially the density of nucleons and helium increases significantly, while some isotopes off the diagonal are basically not populated anymore. This has a direct impact on neutrino production and the quantitative analysis of this effect is ongoing.


\section{The Glashow resonance}

A prominant example where understanding the details of the interaction can make a big difference is the Glashow resonance. A Glashow resonance is a distinct event topology observable in neutrino telescopes, happening when an electron-antineutrino interacts with an electron of the matter in the detector
\begin{equation}
\bar{\nu}_e + e^- \rightarrow W^- \rightarrow \text{ hadrons at 6.3 PeV}.
\end{equation}
The high-energy neutrino, presumably of extragalactic origin, provides the energy to produce a $W^-$-boson, which can decay either hadronically or into leptons. However, in the leptonic scenario, energy is carried out of the detector by neutrinos, such that the reconstruction may not recognize the resonant event. Therefore, we take into account only the hadronic channel here.

Depending on the origin of the incoming neutrino, different event rates can be expected at the detector. This is due to different neutrino flavor compositions at the source from charged pion decays, $\pi^+ \rightarrow \nu_\mu + \mu^+ \rightarrow \nu_\mu + e^+ +\nu_e + \bar{\nu}_\mu$ (conjugated equation for $\pi^-$). For an ideal $p\gamma$-scenario, pions are produced following
\begin{equation}
p + \gamma \rightarrow \Delta^+ \rightarrow \begin{cases} &\pi^+ + n \quad \text{1/3 of all cases} \\ &\pi^0 + p \quad \text{2/3 of all cases} \end{cases}
\end{equation}
and, on the other hand, for an ideal $pp$-scenario
\begin{equation}
p + p \rightarrow \begin{cases} &\pi^+ + \text{anything} \quad \text{1/3 of all cases} \\ &\pi^- + \text{anything} \quad \text{1/3 of all cases} \\ &\pi^0 + \text{anything} \quad \text{1/3 of all cases} \end{cases}.
\end{equation}
Even after flavor mixing, for which we assume tribimaximal mixing, the flavor composition carries the imprint of the production mechanism. The fraction of electron-antineutrinos $\xi_{\bar{\nu}_e}$ at Earth is directly proportional to the expected event number and thus different source scenarios can be distinguished after a certain exposure.

However, as a change of a few percent in $\xi_{\bar{\nu}_e}$ can have a drastic impact on the discrimination power, it is not clear if sources are still distinguishable in a more accurate treatment of the nuclear interactions. For a more realistic description of a $pp$-source, reasonable estimates are obtained from hadronic interaction models like EPOS-LHC \cite{Pierog:2013ria}, QGSJet-II-04 \cite{Ostapchenko:2010vb} and SIBYLL 2.3 \cite{Ahn:2009wx}. We find that the deviation of the pion charge ratio ($\pi^+$ to $\pi^-$) from the ideal scenario for a spectrum of cosmic rays with spectral index $\alpha = 2.0$ is as large as 25\%. This effect increases the softer the cosmic ray spectrum is.

A more realistic $p\gamma$-source takes into account the contamination by $\pi^-$, which are not produced at all in the ideal scenario. These pions can originate from higher resonances in multi-pion processes or from photo-hadronic interactions of neutrons, following
\begin{equation}
n + \gamma \rightarrow \Delta^0 \rightarrow \begin{cases} &\pi^- + p \quad \text{1/3 of all cases} \\ &\pi^0 + n \quad \text{2/3 of all cases} \end{cases}.
\end{equation}
Especially if heavy nuclei are present in the source, many neutrons can be produced as a consequence of nuclear disintegration, which largely enhances $\pi^-$ contamination.

\begin{figure}[t!]
\centering
\includegraphics[width=0.23\textwidth]{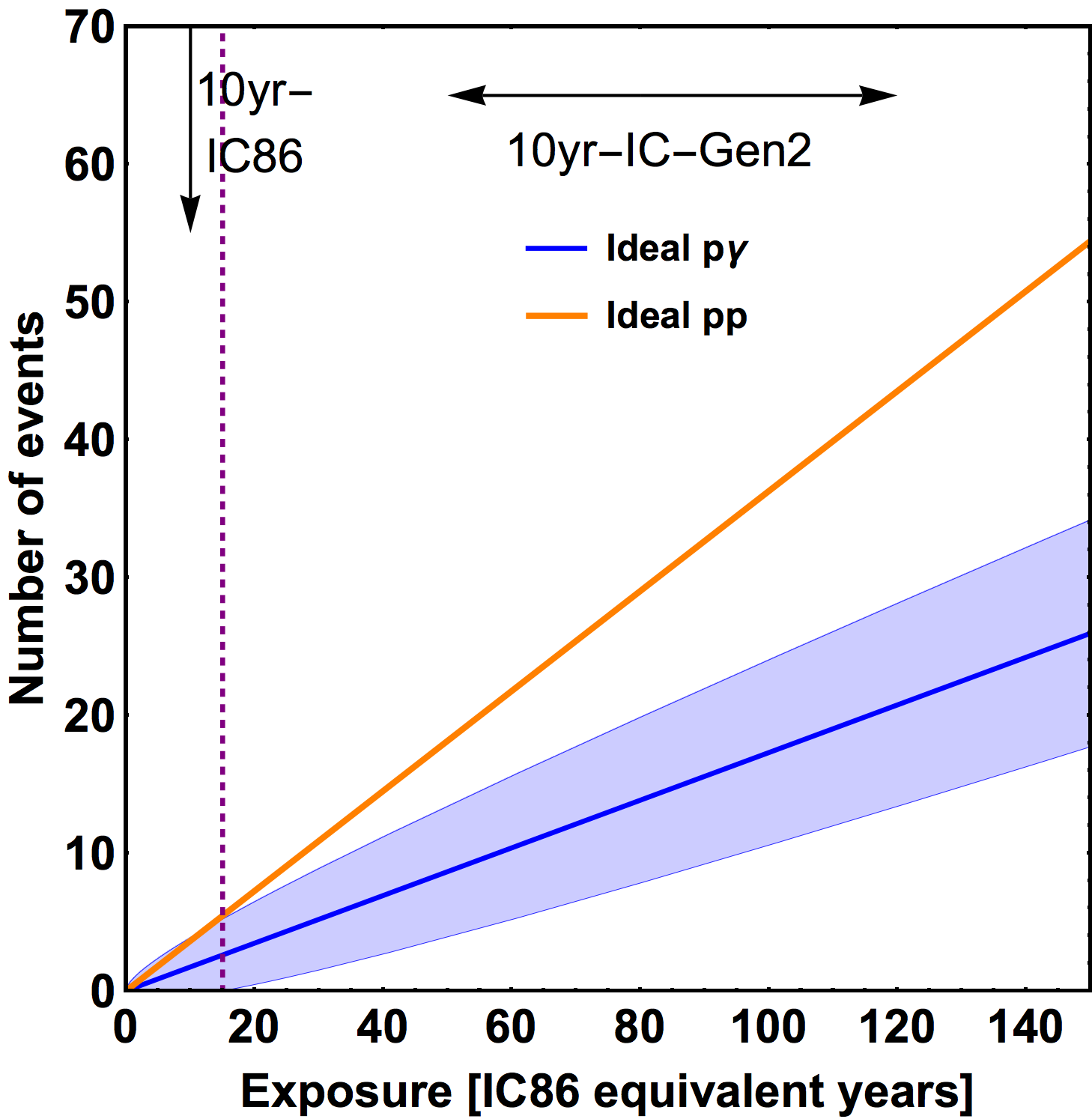}\hspace{0.1cm}
\includegraphics[width=0.23\textwidth]{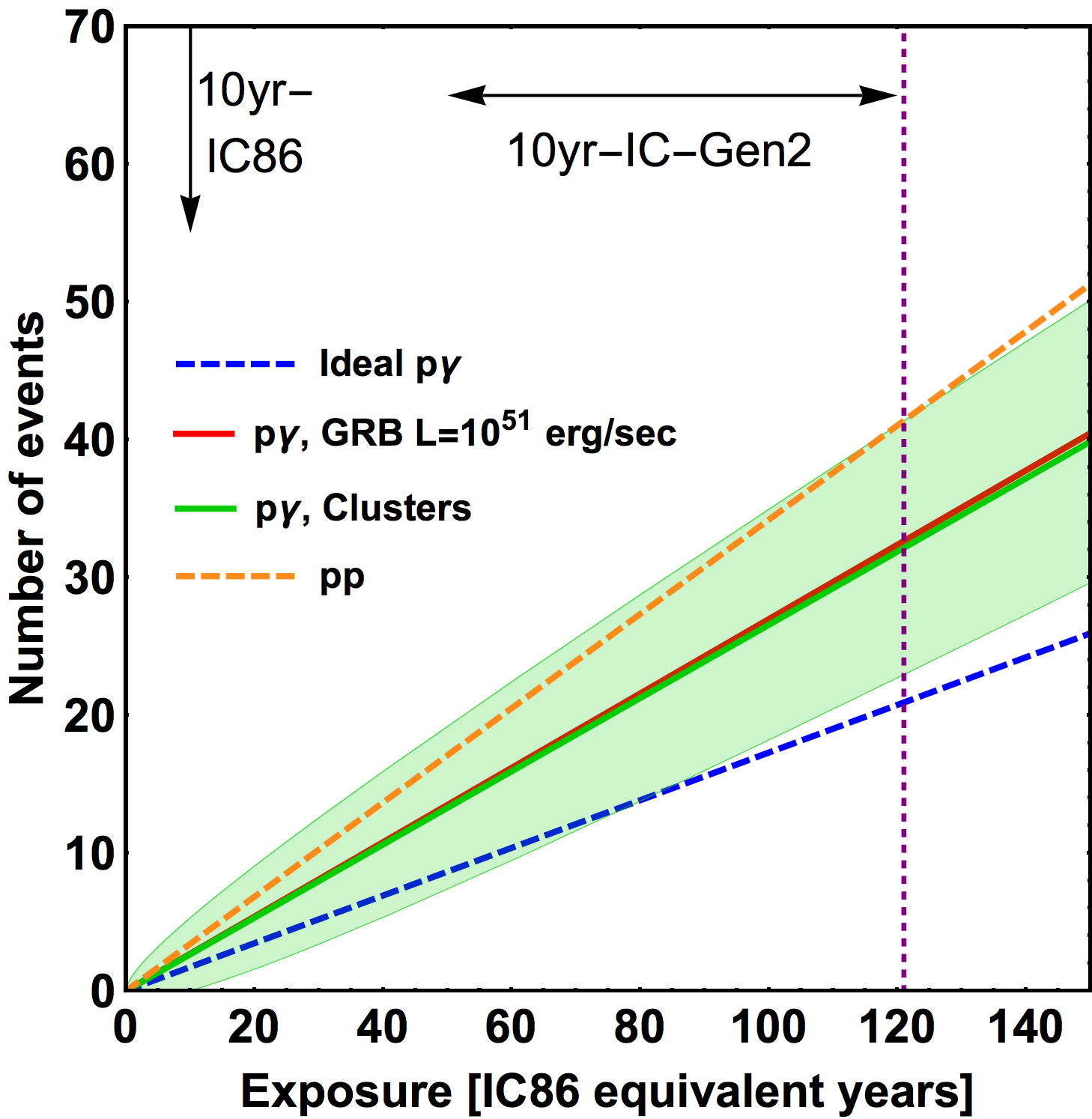}
\caption{Expected number of Glashow resonance events over exposure time in IceCube (86 strings) in the ideal scenario (left) and in a more realistic scenario (right). The equivalent 10 year exposure time in IceCube-Gen2 is indicated by the arrow on the top of each panel. The bands represent statistical (Poissonian) uncertainties and uncertainties of the model (oscillation parameters). The vertical dashed line shows the necessary exposure to discriminate between different source scenarios. Figure taken from \cite{Biehl:2016psj}.}
\label{fig:glashowresonance}
\end{figure}

The effect of these improved descriptions of the interactions is shown in Fig.~\ref{fig:glashowresonance}. The left panel represents the ideal scenarios while on the right hand side the more realistic treatment is illustrated. The colored lines correspond to the expected number of events as a function of exposure time in IceCube (86 strings). We include the statistical (Poissonian) uncertainties as well as the model uncertainties of the oscillation parameters represented by the band attached to the curve (assuming they scale $\propto \sqrt{t}$ with the exposure time). If we have an ideal $p\gamma$-scenario producing neutrinos in the source, it takes about 15 years with IceCube to distinguish it from an ideal $pp$-scenario by the number of resonant events, indicated by the vertical dashed line. If we in turn have a true $pp$-mechanism in the source, a similar picture is obtained for the discrimation from a theoretical $p\gamma$-scenario, which is well within the reach of IceCube-Gen2. The 10 year equivalent exposure is depicted by the arrow in the upper part of the panels.

This discrimination power is lost if the interactions are modelled more realistically. The deviation of the pion charge ratio from one, $\pi^+/\pi^- > 1$, in the $pp$-mechanism and the contamination of the $p\gamma$-source by $\pi^-$ make them look more alike in terms of flavor composition. The two different $p\gamma$-curves correspond to different target photon fields. The red curve assumes a broken power law with spectral indices -1 and -2 and a break at 1 keV similar to GRBs. The green curve corresponds to a synchrotron target photon spectrum from co-accelerated electrons (as in galaxy clusters with $R \approx 10^{19}$ km and $B \approx 10^{-6}$ G). As indicated by the vertical dashed line, we would need about 120 years of IceCube exposure to distinguish both scenarios, which is probably even beyond the reach of Gen2.

However, it may be possible to distinguish sources containing protons only from sources where nuclei interact to produce neutrinos. Depending on the optical thickness of the source to photo-hadronic interactions or the strength of the magnetic field, source properties can be constrained by the Glashow resonance. Even the non-observation of resonant events would exert pressure on certain scenarios like $pp$-, $p\gamma$- or $A\gamma$-sources. On the other hand, it could be interpreted as the effect of muon damping, i.e. muons cool faster than they decay due to strong magnetic fields. 



\section{Summary and conclusions}

We studied astrophysical neutrino production from UHECR interactions in the context of associated uncertainties and how they can change the conclusions drawn from observations.
Prominent source candidates like GRBs, AGNs or TDEs are considered to potentially produce many neutrinos in photo-hadronic interactions on different energy scales.
Most important for neutrino production is the energy range from a few to hundreds of MeV, where nuclear photo-disintegration and photo-meson production occurs.
These processes are mainly responsible for the cosmic ray composition and neutrino production.

We introduced the development of the nuclear cascade as a measure for interactions in a radiation field, where the isotope species are coupled to each other by many competing channels representing energy losses, escape and injection processes. We showed that luminosity and size of the region are the main parameter controlling the development of the nuclear cascade and neutrino production efficiency as they control the radiation density.

However, cross-section data are very sparse as only for a few isotopes on the main diagonal measurements are available. Especially for regions of the parameter space where neutrino production is mildly efficient, this introduces large uncertainties, as neutrinos are mainly produced off heavy nuclei. We showed that interaction models used in astrophysics do not always reproduce the data well such that, depending on the target photons, disintegration rates vary within a factor of 2. Including a systematic bias on partially measured or unmeasured cross-sections, the disintegration chain can significantly alter the development of the nuclear cascade.

We discussed potential extensions for state-of-the-art photo-meson models, such as going beyond the superposition model where photo-hadronic interactions of nuclei are modelled as photons interacting with only a single nucleon. Nuclei tend to show a universal behaviour for photo-meson production where the resonances introduced by the superposition model smear out. Analysis to quantify these effects are ongoing.

The Glashow resonance is a prominent example to stress the importance of understanding the involved interactions. With measuring the neutrino flavor composition we can in principle tell apart different source scenarios.
However, we showed that 
while for the ideal picture it is possible to discriminate $pp$- and $p\gamma$-sources, in a more realistic treatment taking into account the contamination by $\pi^+$ or $\pi^-$, respectively, the discrimination power is lost.


\section*{Acknowledgements}
The authors receive funding from the European Research Council (ERC) under the European Union’s Horizon 2020 research and innovation programme (Grant No. 646623).

\bibliography{references.bib}
%
%
%
%
\vspace{-10cm}
\end{document}